\documentclass[aps,prd,superscriptaddress,twocolumn]{revtex4-1}
\usepackage{graphicx}
\usepackage[caption=false]{subfig}
\usepackage{epstopdf}
\usepackage{amsmath}
\usepackage{amsfonts}
\usepackage{amssymb}
\usepackage{latexsym}
\usepackage{hyperref}
\usepackage[english]{babel}
\usepackage[utf8]{inputenc}
\usepackage[colorinlistoftodos]{todonotes}
\usepackage{color}
\usepackage{slashed}
\usepackage{feynmp}
\usepackage{bm}
\usepackage{bbold}
\usepackage{eufrak}
\usepackage{slashed}
\usepackage{bm}  
\usepackage{tabu}
\usepackage{xcolor}

\begin{document}
\title{Testing Lorentz-symmetry violation via electroweak decays}

\author{Y. M. P. Gomes}\email{ymuller@cbpf.br}
\affiliation{Centro Brasileiro de Pesquisas F\'{i}sicas (CBPF), Rua Dr Xavier Sigaud 150, Urca, Rio de Janeiro, Brazil, CEP 22290-180}

\author{P.C. Malta}\email{pedrocmalta@gmail.com}
\affiliation{Centro Brasileiro de Pesquisas F\'{i}sicas (CBPF), Rua Dr Xavier Sigaud 150, Urca, Rio de Janeiro, Brazil, CEP 22290-180}

\author{M. J. Neves}\email{mneves@ua.edu}
\affiliation{Department of Physics and Astronomy, University of Alabama, Tuscaloosa, Alabama 35487, USA}
\affiliation{Departamento de F\'isica, Universidade Federal Rural do Rio de Janeiro,
BR 465-07, 23890-971, Serop\'edica, RJ, Brazil}



\begin{abstract}
In this work we introduce CPT-odd non-minimal Lorentz-symmetry violating couplings to the electroweak sector modifying the interaction between leptons and gauge bosons. The vertex rules allow us to calculate tree-level processes modified by the presence of the novel dimension-five operators. For definitiveness, we investigate the $W$ decay into a lepton-neutrino pair, the $Z$ decay into pairs of charged and neutral leptons, as well as the decay of the muon. By comparing the experimental measurements on these processes to our results we are able to bound several combinations of the background 4-vectors to be $\lesssim 10^{-4}  \, \mbox{GeV}^{-1}$.
\end{abstract}

\maketitle


\section{Introduction} \label{sec_intro}

\indent

The Standard Model (SM) is based on gauge and Lorentz symmetries and most of its predictions have been experimentally confirmed, including the 2012 discovery of the long-sought Higgs boson. Nonetheless, it is believed that the SM must be the low-energy limit of some broader theory. In many such beyond the SM scenarios, such as string theory and quantum gravity, it is possible that Lorentz symmetry is broken at very high energies as generic tensor fields acquire vacuum expectation values and become time-independent. These are generally coupled to the remaining dynamic physical -- matter, Higgs and gauge -- fields in the low-energy theory that contains the SM, which is no longer Lorentz invariant, since these tensor coefficients select a preferred direction in space-time~\cite{Kost1, Posp1, Mavro1, Mavro2, Moffat, Camelia, ColemanPRD99}. The resulting terms, which are typically suppressed by inverse powers of some large mass or energy scale (e.g., the Planck scale), could generate small physical effects potentially accessible at current or future experiments.

V.A. Kosteleck\'y and D. Colladay have systematically collected the possible low-energy terms arising from Lorentz-symmetry violation (LSV) into the so-called Standard Model Extension (SME)~\cite{Colladay,Colladay2}, which complements the usual SM by introducing novel LSV interactions in all its sectors, from quantum chromodynamics to gravitation. Diverse experimental tests, ranging from atomic spectroscopy to astrophysical observations passing by collider experiments, have placed bounds on many of the possible LSV coefficients, which are conveniently collected in ref.~\cite{tables} (see also ref.~\cite{Mattingly}).

Experimental tests of standard quantum electrodynamics (QED) are extremely precise, allowing the corresponding sector in the SME to be well constrained. A prime example is the Chern-Simons-like, CPT-odd Carroll-Field-Jackiw ($d = 3$) correction to the photon sector~\cite{CFJ}, which is tightly constrained by the non-observation of the rotation in the polarization of radiation from astrophysical sources~\cite{Mewes1, Mewes2}. Though complementary, laboratory-based tests are not entirely competitive~\cite{Pedro_Yuri}. The CPT-even sector ($d = 4$), on the other hand, can be strongly bounded by a variety of laboratory experiments and by astrophysical observations such as gravitational-wave detectors~\cite{ligo, tables}. Both sectors are renormalizable and are at most quadratic in the photon fields. Non-renormalizable, higher-derivative terms may also be introduced, but are expected to be suppressed relative to their renormalizable counterparts~\cite{highdev, Borges}.

The electroweak sector of the SME, on the other hand, has not been studied to the same extent. In the SM electroweak processes are generally harder to detect than pure electromagnetic ones due to the presence of inverse powers of the large mass of the mediating bosons in the amplitudes. In well-measured processes, such as Bhabha scattering~\cite{Bhabha1, Bhabha2}, QED effects are the leading contributions, whereas electroweak effects amount to only a few percent at energies already close to the $Z$ pole~\cite{Bhabha3, scat_LSV}. Since at low energies $W$- and $Z$-mediated processes are strongly suppressed relative to photon-mediated ones, we are going to focus on purely electroweak interactions.

Lorentz-symmetry violation may be incorporated into the electroweak sector by considering couplings analogous to those in the QED sector of the (minimal) SME. A possibility is to introduce a Chern-Simons-like operator which generalizes the Carroll-Field-Jackiw case, but with the interesting feature of a term coupling the photon and $Z$ boson, which leads to photon-$Z$ mixing~\cite{Potting_1}. Another interesting possibility is to modify the propagator of the intermediate bosons, thus affecting any $W$- or $Z$-mediated processes such as muon decay~\cite{Onderwater}, neutron $\beta$ decay~\cite{Vos_1} and nuclear processes~\cite{Noordmans, Dijck, Ullman}. There is however another interesting possibility, namely to directly modify the interaction vertices between mediators and matter fields.

In QED and, more generally, in the SM, gauge fields are introduced as the result of the invariance of the fermionic Lagrangian under local phase transformations. These fields are coupled to matter in such a way that the usual partial derivative in the fermionic kinetic terms may be generalized into a covariant derivative {\it \`a la} $\partial_\mu \rightarrow D_\mu = \partial_\mu + iq A_\mu$, where $q$ is a (conserved) charge and $A_\mu$ is the gauge field associated to the -- here Abelian -- symmetry. The ensuing coupling between the gauge field and matter fields is called minimal, but it is not the only possibility.

If a premium is placed on gauge invariance, an interesting possibility is the introduction of non-minimal couplings, which consist of coupling the fermionic bilinears not to the gauge fields themselves, but to their field-strength tensors as in the Pauli term~\cite{Pauli}. Non-minimal couplings have been explored in many different contexts within Lorentz-violating QED, such as the spectrum of the hydrogen atom~\cite{Belich1}, magnetic and electric dipole moments of charged leptons~\cite{Posp3, Posp4, Belich2}, scattering processes~\cite{Maluf, scat_LSV}, and topological effects~\cite{Bakke, Anacleto}.

In this paper, we generalize the couplings discussed above and introduce LSV terms directly coupling the leptonic bilinears with the field-strength tensors of the non-Abelian gauge bosons, thereby extending the lepton-gauge interactions beyond the usual, Lorentz-preserving minimal couplings from the $SU_{\rm L}(2)\times U(1)_{\rm Y}$ symmetry of the SM. These novel terms produce modifications already at tree level and in this context we shall address the effects of LSV in $W$ decay $W^{-} \, \rightarrow \, \overline{\nu}_\ell \, \ell$, the $Z$ decay $Z \, \rightarrow \, \overline{f} \, f$, where $f$ is any SM lepton or neutrino, as well as to muon decay, a purely leptonic process of historical and practical importance in tests of the SM~\cite{Renga}. By using the data presented in the latest edition of the Particle Data Group~\cite{PDG}, we are able to constrain different combinations of the LSV coefficients.


This paper is organized as follows: in section~\ref{sec_model_lg} we present the novel LSV terms and explicitly construct the Lagrangian for the lepton-gauge sector with LSV interactions. In section~\ref{sec_apps} we apply the LSV-modified Feynman rules to a few processes at tree level to obtain upper bounds on the LSV parameters. Finally, in section~\ref{sec_conclusion} we summarize our results and present our concluding remarks. In our calculations we employed the Package-X~\cite{PackageX} to automatically evaluate the traces and contractions involving spinors and Dirac gamma matrices. We use natural units ($c = \hbar = 1$) throughout.

\section{The LSV Lepton-gauge interactions} \label{sec_model_lg}
\indent

In the SM, the leptons are minimally coupled to the gauge bosons via the covariant derivative
\begin{equation} \label{dmu}
D_\mu \equiv \partial_\mu - i g' \, Y \, B_\mu - ig \, W_\mu^a \, \sigma^a/2 \; ,
\end{equation}
where $g'$ and $g$ are the respective $U(1)_{\rm Y}$ and $SU_{\rm L}(2)$ coupling constants, $Y$ is the weak hypercharge and $\left\{ \sigma^{a} \right\}$ are the Pauli matrices. The tree-level interactions between leptons $\psi$ and gauge bosons in the SM are exclusively derived from terms $\sim i\bar{\psi}\gamma^\mu D_\mu \psi$.

Here we introduce LSV non-minimal couplings, i.e., interaction terms between the leptons and the field-strength tensors of the $U(1)_{\rm Y}$ and $SU_{L}(\rm 2)$ gauge fields, $B_{\mu\nu} = \partial_\mu B_\nu - \partial_\nu B_\mu$ and $W_{\mu\nu}^{a} = \partial_\mu W_\nu^a - \partial_\nu W_\mu^a + g \, \varepsilon^{abc} \, W_\mu^b \, W_\nu^c$ $(a=1,2,3)$, respectively. This is accomplished via two real (constant) 4-vectors $\xi^{\mu}$ and $\rho^{\mu}$ that give rise to the following LSV lepton-gauge Lagrangian
\begin{eqnarray}
\mathcal{L}_{\rm g\ell}^{\rm LSV} & = & -\overline{\psi}_{\rm \ell L} \, \gamma^\mu \left( \frac{1}{2} \, \xi^\nu \, B_{\mu \nu} + \rho^\nu \, W^a_{\mu \nu} \, \frac{\sigma^a}{2} \right) \psi_{\rm \ell L}  \nonumber \\
& - & \overline{\ell}_{\rm R} \, \gamma^\mu \left( \frac{1}{2} \, \xi^\nu \, B_{\mu \nu} \right) \ell_{\rm R} \; ,\label{eq_lag_lsv}
\end{eqnarray}
which introduces interactions with extra momentum-dependent contributions -- a typical signature of such LSV couplings~\cite{Maluf, scat_LSV}. We would like to mention that similar couplings have been explored in refs.~\cite{UFMA, erratumUFMA}. Here, however, we conduct a more general analysis by considering the $SU_{\rm L}(2)$ and $U(1)_{\rm Y}$ sectors together.

A few comments are in order. The internal symmetries of the theory remain untouched: the charge operator is given by $Q_{em} = Y + I^3$, where $I^{3}=\sigma^{3}/2$, so that the charge assignments of the matter fields are the same as in the SM, namely $\psi_{\rm \ell L} = \left( \, \, \nu_{\ell} \; \; \; \ell \, \, \right)^T_{\rm L} \,\sim\,  \left( \, {\bf 2}, -1/2 \, \right)$ and $\ell_{\rm R} \,\sim\, \left( \, {\bf 1}, -1 \, \right)$, in which $\ell = \left\{ \, e \, , \, \mu \, , \, \tau \, \right\}$. Here, $\psi_{\rm R,L} = P_{\rm R,L}\psi$, where $P_{\rm R,L} \equiv \frac{1}{2}\left( 1 \pm \gamma_5  \right)$ are the right- and left-handed projection operators. Right-handed neutrinos are singlets under $SU_{\rm L}(2) \times U(1)_{\rm Y}$ and are {\it a priori} not contained in the SM apart from issues related to neutrino masses and mixing, which are not going to be of consequence here, since we treat neutrinos as massless.

It is important to note that the LSV 4-vectors above may distinguish between lepton families. We tacitly assume that the process leading to the breaking of Lorentz symmetry equally affects the gauge sectors of the $SU_{\rm L}(2)$ and $U(1)_{\rm Y}$ symmetries. However, there is no reason to suppose that the matter fields are indiscriminately affected, meaning that the LSV 4-vectors are actually family-dependent parameters: $\xi \rightarrow \xi_{(\ell)}$ and $\rho \rightarrow \rho_{(\ell)}$. To avoid overloading the notation, we shall specialize to each family only at the end of the respective calculation, where different experimental limits may apply.

The interaction terms from~$D_\mu$, eq.~\eqref{dmu}, explicitly depend on the hypercharge and isospin of the leptons. The LSV terms in eq.~\eqref{eq_lag_lsv}, however, are insensitive to those~\cite{Belich1, KLZ, Belich3}. The mass terms from the Yukawa interactions are not affected, since the LSV terms do not influence them at tree level, meaning that the equations of motion for the free leptons are unchanged and the propagators -- and associated Feynman rules -- will be the same as in the SM.

At last but not least, we would like to mention that the LSV couplings have negative canonical dimension, i.e., the LSV terms are non-renormalizable. This is a general feature of such non-minimal couplings and indicates that the associated Lagrangian is, in fact, only an effective theory. This will not disturb us here, since we are only dealing with relatively low-energy processes -- exclusively at tree level --, so no divergences are expected.

We are now able to dissect eq.~\eqref{eq_lag_lsv} further and determine the Feynman rules governing the lepton-boson interactions. We may decompose the lepton-gauge Lagrangian into $\mathcal{L}_{\rm g\ell} = \mathcal{L}_{\rm g\ell}^{\rm SM} + \mathcal{L}_{\rm g\ell}^{\rm LSV}$, where the first term contains the SM contributions to processes involving neutral and charged currents and the second contains only LSV terms from eq.~\eqref{eq_lag_lsv}.

The mechanism of spontaneous symmetry breaking is the same as in the SM, so we apply the standard Weinberg rotation to write $B_\mu$ and $W^3_\mu$ in terms of $A_\mu$ and $Z_\mu$~\cite{Mandl}. The physical fields $W^{\pm}$ and $Z$ have masses $m_{W}=80$~GeV and $m_{Z}=91$~GeV, respectively, and $A$ represents the massless photon. The field-strength tensors of the photon and the $Z$ boson are defined as $F_{\mu\nu}=\partial_{\mu}A_{\nu} - \partial_{\nu}A_{\mu}$ and $Z_{\mu\nu}=\partial_{\mu}Z_{\nu} - \partial_{\nu}Z_{\mu}$, while $W_{\mu\nu}^{3}$ and $F_{\mu\nu}^{+}$ (with $F_{\mu\nu}^{-} $ being its complex conjugate and $W^{\pm}_{\mu\nu} \equiv \partial_{\mu} W_{\nu}^{\pm} - \partial_{\nu} W_{\mu}^{\pm}$) are given by
\begin{eqnarray}
W_{\mu\nu}^{3} & = & \cos\theta_{W} Z_{\mu\nu} + \sin\theta_{W} F_{\mu\nu} \nonumber \\
& - & i g \left( W_{\mu}^{+}  W_{\nu}^{-}-W_{\nu}^{+} W_{\mu}^{-} \right) \; ,
\\
F_{\mu\nu}^{+} & = & W^{+}_{\mu\nu} +  i g \cos\theta_{W} \left( W_{\mu}^{+} Z_{\nu} - Z_{\mu} W_{\nu}^{+}  \right) \nonumber \\
& + & i g \sin\theta_{W} \left( W_{\mu}^{+} A_{\nu} - W_{\nu}^{+} A_{\mu} \right) \; .
\end{eqnarray}

With the relations above we write the LSV piece as
\begin{eqnarray}\label{eq_lag_2}
\mathcal{L}_{\rm g\ell}^{\rm LSV} & = & \frac{1}{2} \, \cos\theta_W \, \xi^\mu  \left( \bar{\psi}_{\rm \ell L} \gamma^\nu \psi_{\rm \ell L} + \bar{\ell}_{\rm R} \gamma^\nu \ell_{\rm R} \right)F_{\mu\nu}  \\
& - & \frac{1}{2} \, \sin \theta_W \, \xi^\mu \left( \bar{\psi}_{\rm \ell L} \gamma^\nu \psi_{\rm \ell L} + \bar{\ell}_{\rm R} \gamma^\nu \ell_{\rm R} \right) Z_{\mu\nu}  \nonumber \\
& + & \rho^\mu \overline{\psi}_{\rm \ell L} \gamma^\nu \left( F_{\mu\nu}^{+}\frac{\sigma^{+}}{2}+F_{\mu\nu}^{-}\frac{\sigma^{-}}{2} + W_{\mu\nu}^{3}\frac{\sigma^{3}}{2}  \right) \psi_{\rm \ell L} \; , \nonumber
\end{eqnarray}
where $\sqrt{2} \, \sigma^{\pm}=\sigma^{1} \pm i \, \sigma^{2}$. The Lagrangian involving only left-handed leptons reads then
\begin{eqnarray}\label{eq_L}
\mathcal{L}_{\rm g\ell, L}^{\rm LSV} & = &  \frac{1}{2} v_1^\mu \overline{\ell}_{L} \gamma^\nu \ell_{L} F_{\mu\nu} + \frac{1}{2} v_2^\mu  \overline{\nu}_{\ell L} \gamma^\nu \nu_{\ell L} F_{\mu\nu}  \nonumber \\
& + & \frac{1}{2}  v_3^\mu  \overline{\ell}_{L} \gamma^\nu \ell_{L} Z_{\mu\nu} +\frac{1}{2}  v_4^\mu  \overline{\nu}_{\ell L} \gamma^\nu \nu_{\ell L} Z_{\mu\nu}  \nonumber \\
& + & \frac{i g}{2} \, \rho^\mu  \left( \bar{\ell}_{L} \gamma^\nu \ell_{L} - \bar{\nu}_{\ell L} \gamma^\nu \nu_{\ell L} \right)W^+_{[\mu} \, W^-_{\nu]}   \nonumber \\
& + & \frac{1}{\sqrt{2}}  \rho^\mu \overline{\ell}_{L} \gamma^\nu \nu_{\ell L} \Big( W^{-}_{\mu\nu} -i e A_{[\mu}W_{\nu]}^-   \nonumber \\
& - & i e \cot \theta_W Z_{[\mu}W_{\nu]}^- \Big) + {\rm H. c.} \, ,
\end{eqnarray}
where we defined $A_{[\mu}B_{\nu]} \equiv A_{\mu}B_{\nu} - B_{\mu}A_{\nu}$. The coupling constants $g$ and $g^{\prime}$ are connected via $e = g\sin\theta_W = g'\cos\theta_W$, where $e \simeq \sqrt{4\pi/128}\simeq 0.31$ is the fundamental electric charge and $\theta_{W}$ is the Weinberg angle satisfying $\sin^{2}\theta_{W}=0.23$~\cite{PDG}. For simplicity, we have defined the rotated vectors
\begin{subequations}
\begin{eqnarray}
v_{1\mu} &=& \cos \theta_W \xi_{\mu} - \sin \theta_W \rho_{\mu} \; ,\label{v1}
\\
v_{2\mu} &=&  \cos \theta_W \xi_{\mu} + \sin \theta_W \rho_{\mu} \; ,
\\
v_{3\mu} &= & -\sin \theta_W \xi_{\mu} - \cos \theta_W \rho_{\mu} \; ,
 \\
v_{4\mu} &=&  -\sin \theta_W \xi_{\mu} + \cos \theta_W \rho_{\mu} \; .\label{v4}
\end{eqnarray}
\end{subequations}

\setlength{\tabcolsep}{4pt}
\renewcommand{\arraystretch}{0.9}
\newcolumntype{C}[1]{>{\centering\arraybackslash}m{#1}}
\begin{table}[]
\centering
\begin{tabular}[t]{@{}|C{2cm}|C{3.5cm}|C{-0.6cm}@{}}
\cline{1-2}
interaction  & vertex factor   &   \\  [5pt]
\cline{1-2}
$\gamma \, \ell \, \bar{\ell}$  &   $( c_1^{[\mu} \gamma^{\nu]} + c_2^{[\mu} \gamma^{\nu]}\gamma_5 ) q_\nu $    &   \\ [10pt]
\cline{1-2}
$Z^0 \, \ell \, \bar{\ell}$ &  $( c_3^{[\mu} \gamma^{\nu]} + c_4^{[\mu} \gamma^{\nu]}\gamma_5 ) q_\nu$    &   \\ [10pt]
\cline{1-2}
$Z^{0} \, \nu_{\ell} \, \bar{\nu}_{\ell}$ & $\frac{1}{4} \, v_4^{[\nu} \gamma^{\mu]} \left(1-\gamma_5\right) \, q_\nu$    &   \\ [10pt]
\cline{1-2}
$W^{-} \, \ell \, \bar{\nu}_{\ell}$ & $\frac{1}{2\sqrt{2}} \, \rho^{[\nu} \gamma^{\mu]} \left(1-\gamma_5\right) \, q_\nu$    &   \\ [10pt]
\cline{1-2}
\end{tabular}
\captionsetup{justification=centering}
\caption{List of vertices that are present at tree level in the SM and receive a (small) LSV correction from eq.~\eqref{eqfinal}. Here, $q^\mu$ is the 4-momentum of the photon, $W$ or $Z$ boson flowing into the vertex. The coefficients $v_{i}^{\mu}$ and $c_{i}^{\mu}$ are listed in eqs.~\eqref{v1}-\eqref{v4} and in eqs.~\eqref{c1}-\eqref{c4}, respectively.}
\label{table:list_vert_sm}
\end{table}

\setlength{\tabcolsep}{4pt}
\renewcommand{\arraystretch}{0.9}
\newcolumntype{C}[1]{>{\centering\arraybackslash}m{#1}}
\begin{table}[]
\centering
\begin{tabular}[t]{@{}|C{2cm}|C{3.5cm}|C{-0.6cm}@{}}
\cline{1-2}
interaction  & vertex factor   &   \\  [5pt]
\cline{1-2}
$\gamma \, \nu_{\ell} \, \bar{\nu}_{\ell}$  &  $\frac{1}{4} \, v_2^{[\nu} \gamma^{\mu]} \left(1-\gamma_5\right) \, q_\nu$    &   \\ [10pt]
\cline{1-2}
$W^{-} \, \gamma \, \ell \, \bar{\nu}_{\ell}$ & $-\frac{i e}{2\sqrt{2}} \, \rho^{[\nu} \gamma^{\mu]} \left(1-\gamma_5\right)$    &   \\ [10pt]
\cline{1-2}
$W^{-} \, Z^0 \, \ell \, \bar{\nu}_{\ell}$ & $-\frac{i e \, \cot\theta_W }{2\sqrt{2}} \rho^{[\nu} \gamma^{\mu]} \left(1-\gamma_5\right)$    &   \\ [10pt]
\cline{1-2}
$W^{+} \, W^{-} \, \ell \, \bar{\ell}$  & $\frac{i g}{4} \, \rho^{[\nu} \gamma^{\mu]} \left(1-\gamma_5\right)$    &   \\ [10pt]
\cline{1-2}
$W^{+} \, W^{-} \, \nu_{\ell} \, \bar{\nu}_{\ell}$ & $-\frac{i g}{4} \, \rho^{[\nu} \gamma^{\mu]} \left(1-\gamma_5\right)$    &   \\ [10pt]
\cline{1-2}
\end{tabular}
\captionsetup{justification=centering}
\caption{List of novel vertices from eq.~\eqref{eqfinal} that are in principle absent from the SM at tree level. Here, $q^\mu$ is the 4-momentum of the photon, $W$ or $Z$ boson flowing into the vertex. The coefficients $v_{i}^{\mu}$ and $c_{i}^{\mu}$ are listed in eqs.~\eqref{v1}-\eqref{v4} and in eqs.~\eqref{c1}-\eqref{c4}, respectively.}
\label{table:list_vert_lsv}
\end{table}

Let us now return to the full LSV interaction Lagrangian. Including the results from eq.~\eqref{eq_L} and using the definition of the left- and right-handed projectors, we are able to rewrite eq.~\eqref{eq_lag_2} in terms of the basic lepton fields with the usual V-A vertex structure of the electroweak theory. The result is
\begin{eqnarray}\label{eqfinal}
\mathcal{L}_{\rm g\ell}^{\rm LSV} & = &  \overline{\ell} \left( c_1^\mu \gamma^\nu + c_2^\mu \gamma^\nu \gamma_5  \right) \ell \, F_{\mu\nu}  \\
& + & \frac{1}{4} \, v_2^\mu \, \overline{\nu}_{\ell}  \, \gamma^\nu \left(1-\gamma_5\right) \nu_{\ell} \, F_{\mu\nu} \nonumber \\
& + & \overline{\ell} \left( \, c_3^\mu \, \gamma^\nu + c_4^\mu \, \gamma^\nu \, \gamma_5 \, \right) \ell \, Z_{\mu\nu}
\nonumber \\
& + & \frac{1}{4} \, v_4^\mu \, \overline{\nu}_{\ell} \, \gamma^\nu \left(1-\gamma_5\right) \nu_{\ell} \, Z_{\mu\nu}  \nonumber \\
& + & \frac{ig}{4}  \rho^\mu  W^+_{[\mu} W^-_{\nu]} \left[ \overline{\ell} \, \gamma^\nu \left(1-\gamma_5\right) \ell
- \overline{\nu}_{\ell} \gamma^\nu \left(1-\gamma_5\right) \nu_{\ell} \right]
\nonumber \\
& + & \frac{1}{2\sqrt{2}}  \rho^\mu \overline{\ell} \gamma^\nu \left(1-\gamma_5\right) \nu_{\ell} \times \nonumber \\
& \times & \left( W^{-}_{\mu\nu} -i e A_{[\mu}W_{\nu]}^-
- i e \cot \theta_W Z_{[\mu}W_{\nu]}^{-} \right)
+ \mbox{H.c.}, \nonumber
\end{eqnarray}
where the coefficients $c_{i}^{\mu}$are given by
\begin{subequations}
\begin{eqnarray}
c_{1\mu} &=& \frac{1}{2}\left(\cos \theta_W \xi_{\mu} - \frac{1}{2}\sin \theta_W \rho_{\mu} \right)
\; , \hspace{0.3cm}\label{c1}
\\
c_{2\mu} &=& \frac{1}{4}\sin \theta_W \rho_{\mu} \; ,
\hspace{0.3cm}
\\
c_{3\mu} &=& -\frac{1}{2}\left(\sin \theta_W \xi_{\mu} + \frac{1}{2}\cos \theta_W \rho_{\mu}  \right) \; ,
\hspace{0.3cm}
\\
c_{4\mu} &=& \frac{1}{4}\cos \theta_W \rho_{\mu} \; .
\hspace{0.3cm}\label{c4}
\end{eqnarray}
\end{subequations}

Equation~\eqref{eqfinal} is our final result and shows how eq.~\eqref{eq_lag_lsv} modifies the usual lepton-boson interactions. Several vertices from the SM receive small LSV corrections and other new purely LSV vertices are introduced. Now we may extract the interaction vertices to compute physical observables (cf. tables~\ref{table:list_vert_sm} and~\ref{table:list_vert_lsv}.).

It is worthwhile mentioning that some of the interaction terms displayed in eq.~\eqref{eqfinal}, in particular the first one that concerns charged leptons and the photon field-strength tensor, have already been defined in the SME~\cite{tables} (see Table P58). These are analogous to $a_{\rm F}^{(5) \mu\alpha\beta}$ and $b_{\rm F}^{(5) \mu\alpha\beta}$, which couple the vector and pseudo-vector matter currents to the electromagnetic field-strength tensor, respectively. In appendix~\ref{ap_a_F} we discuss the connection between our LSV coefficients and the ones listed in ref.~\cite{tables} in more detail.

Contrary to other, unrelated dimension-5 couplings, like $a_{\rm eff}$, $\hat{a}$ and $\mathring{a}$ (and corresponding $b$ versions)~\cite{Kost_fermion1, Kost_fermion2}, these coefficients have not been extensively constrained in the literature. Indeed, the only bound available is $< 10^{-3} \; {\rm GeV}^{-1}$ from Bhabha scattering (see ref.~\cite{scat_LSV} and Table D22 in ref.~\cite{tables}). In what follows we will improve this limit by up to one order of magnitude.

\section{Application to selected electroweak processes} \label{sec_apps}
\indent

In the previous section we developed the LSV Lagrangian and attained eq.~\eqref{eqfinal}, from where we may read the Feynman rules for the vertices as listed in tables~\ref{table:list_vert_sm} and~\ref{table:list_vert_lsv}. Our goal is to calculate observable quantities, in particular decay widths, that have been experimentally measured and, under the -- so far justified -- assumption that the SM appropriately describes the central value of the experimental results, use the quoted uncertainties to extract upper limits on the LSV coefficients.


\subsection{The $W$ decay width} \label{sec_W_decay}
\indent

Let us first consider the decay of the $W^{-}$ boson into a lepton and its anti-neutrino. The $W^{-}$ boson starts with 4-momentum $k^{\mu}$ and polarization vector $\epsilon_{\mu}(k)$, whereas the decay products have 4-momenta $q^{\mu}$ (lepton) and $q^{\prime \mu}$ (anti-neutrino). The tree-level amplitude is
\begin{equation}\label{amp_W}
i M(W^{-} \rightarrow \ell \, \overline{\nu}_{\ell}) = \epsilon_{\mu}(k) \, \overline{u}_{\ell}(q) V^{\mu}_{W\ell\bar{\nu}_\ell}(k) v_{\bar{\nu}}(q') \; ,
\end{equation}
where $\overline{u}_{\ell}$ and $v_{\bar{\nu}}$ are the Dirac spinors for the lepton and anti-neutrino, respectively. The relevant vertex, including the charged-current interaction from the SM and the LSV contribution (cf. table~\ref{table:list_vert_sm}), is
\begin{eqnarray}
V^{\mu}_{W\ell\bar{\nu}_\ell}(k) & = & -\frac{ig}{2\sqrt{2}} \, \gamma^{\mu}(1-\gamma_{5}) \nonumber \\
& + &  \frac{1}{2\sqrt{2}} \left(\rho^{\nu}\gamma^{\mu}-\rho^{\mu}\gamma^{\nu} \right)(1-\gamma_{5}) k_{\nu} \; .   \label{vertex_W}
\end{eqnarray}

Since we are interested in the unpolarized decay rate, we need to average the squared amplitude over initial polarizations and sum over final spins. Using the fact that $\left( 1 - \gamma_5  \right) \gamma^\mu \left( 1 + \gamma_5  \right) = 2 \gamma^\mu \left( 1 + \gamma_5  \right)$, the spin-averaged square amplitude is given by
\begin{eqnarray}
\langle |M|^2  \rangle & = & \frac{1}{12} \left(-\eta_{\mu\lambda} + \frac{k_\mu k_\lambda}{m_W^2}  \right) q'_\alpha q_\beta  \nonumber \\
& \times &  {\rm Tr}\left[ \overline{\Gamma}^\mu_{+} \gamma^\alpha \left(1 + \gamma_5 \right) \overline{\Gamma}^\lambda_{-} \gamma^\beta \right] \; , \label{amp_W_1}
\end{eqnarray}
where the $\overline{\Gamma}^\mu_{\pm}$ matrices are defined by
\begin{eqnarray}\label{v_operator}
\overline{\Gamma}^\mu_{\pm} & = &  g \gamma^{\mu} \pm i \left(\rho^{\nu}\gamma^{\mu} - \rho^{\mu}\gamma^{\nu} \right) k_{\nu} \; .
\end{eqnarray}

It is now convenient to move to the rest frame of the $W^{-}$ boson, where $k^{\mu}=(m_{W},{\bf 0})$. Since $m_W \gg m_\ell, \; m_{\nu_\ell}$, we may ignore the smaller masses, so that eq.~\eqref{amp_W_1} is the sum of the following partial amplitudes:
\begin{eqnarray}
\langle |M|^2  \rangle_{\rm SM} & = & \frac{g^2}{3m_W^2} \left[ 2\left(k\cdot q\right)\left(k\cdot q^{\prime}\right) + m_W^2 q\cdot q^{\prime}  \right] \; , \label{amp_W_SM}  \\
\langle |M|^2  \rangle_{\rm LSV}^{(1)} & = &  -\frac{2g}{3} k^\mu q^{\prime \nu} q^\alpha \rho^\beta \epsilon_{\mu\nu\alpha\beta} \; ,  \label{amp_W_LSV_1}  \\
\langle |M|^2  \rangle_{\rm LSV}^{(2)} & = & \frac{2m_W \rho_0}{3} \bigg\{ \left( k\cdot q \right)\left( \rho\cdot q^{\prime} \right) + \left( k\cdot q^{\prime} \right)\left( \rho\cdot q \right) +
\nonumber \\
&&
\hspace{-0.8cm}
- \frac{\rho^2}{2m_W \rho_0} \left[ 2\left(k\cdot q\right)\left(k\cdot q^{\prime}\right) - m_W^2 q\cdot q^{\prime}  \right] \bigg\} \; .   \label{amp_W_LSV_2}
\end{eqnarray}

In the rest frame of the $W^{-}$ boson we may use momentum conservation to show that $q^\mu = \frac{m_W}{2}\left(1, \hat{{\bf u}} \right)$ and $q^{\prime \mu} = \frac{m_W}{2}\left(1, -\hat{{\bf u}} \right)$, where $\hat{{\bf u}}$ is a unitary vector in the direction of the 3-momentum of the outgoing lepton. Furthermore,  at the vertex we have $k = q + q^\prime$, which makes eq.~\eqref{amp_W_LSV_1} identically zero. Incorporating all this in the equations above finally gives us
\begin{equation}\label{amp_W_2}
\langle |M|^2  \rangle = \frac{g^2 m_W^2}{3} \left( 1 + \frac{m_W^2 \rho_0^2}{g^2}  \right) \; ,
\end{equation}
which shows no first-order LSV contribution  It is also worthwhile noticing that the LSV piece depends only on the isotropic time component $\rho_0$.

The general expression for the unpolarized two-body decay rate of the $W^{-}$ boson is
\begin{eqnarray}
\Gamma(W^{-} \rightarrow \ell \; \overline{\nu}_{\ell}) & = & \frac{1}{32\pi^2 m_W} \int \frac{d^3{\bf q} \; d^3{\bf q}^\prime }{E_\ell E_{\bar{\nu}_\ell}}  \nonumber \\
& \times & \langle|M|^{2}\rangle \, \delta^{(4)}\left(k- q-q^{\prime}\right) \; ,\label{Gamma}
\end{eqnarray}
and, given that both the SM and LSV contributions contain no angular factors, we are able to perform the phase-space integrals in the same way as in the SM. Dividing this by the full width $\Gamma_W$ gives us the branching ratio for the channel $W^{-} \rightarrow \ell \; \overline{\nu}_{\ell}$, so that, using $G_{F} = \sqrt{2}g^2/8m_W^2$, we have
\begin{equation}\label{gamma_W}
{\rm BR}(W^{-} \rightarrow \ell \; \overline{\nu}_{\ell}) = \frac{G_{F} \; m_{W}^{3}}{6\sqrt{2}\pi \; \Gamma_W} \left(1 +\frac{\rho_0^2}{4 \sqrt{2} \; G_F } \right).
\end{equation}

The first term in eq.~\eqref{gamma_W} is the well-known result from the SM, whereas the second is a small deviation arising from the couplings introduced in eq.~\eqref{eq_lag_lsv}. As mentioned in the introduction, the LSV couplings are actually family dependent, so we may use the experimentally measured branching ratios for each channel~\cite{PDG} 
\begin{subequations}
\begin{eqnarray}
{\rm BR}(W^{-} \rightarrow e^{-} \; \overline{\nu}_{e})_{\rm exp} & = & \left( 10.71 \pm 0.16 \right) \, \% \; ,
\label{exp_lim_W_e} \\
{\rm BR}(W^{-} \rightarrow \mu^{-} \; \overline{\nu}_{\mu})_{\rm exp} & = & \left( 10.63 \pm 0.15 \right) \, \% \; ,
\label{exp_lim_W_mu} \\
{\rm BR}(W^{-} \rightarrow \tau^{-} \; \overline{\nu}_{\tau})_{\rm exp} & = & \left( 11.38 \pm 0.21 \right) \, \% \; ,
\label{exp_lim_W_tau}
\end{eqnarray}
\end{subequations}
to constrain our LSV coefficients.

The results above are well fitted by the SM, given by the first term in eq.~\eqref{gamma_W}, so we may assume that the LSV effects are hidden within the experimental uncertainties and demand the second term in eq.~\eqref{gamma_W} to be smaller that the relative experimental errors in eqs.~\eqref{exp_lim_W_e}-\eqref{exp_lim_W_tau}. By doing so we obtain the upper bounds (at $1 \, \sigma$)
\begin{subequations}
\begin{eqnarray}
|\rho_{(e)}^0| &\lesssim & 9.8 \times 10^{-4} \, \mbox{GeV}^{-1} \; ,
\label{W_bound_e} \\
|\rho_{(\mu)}^0| &\lesssim & 9.6 \times 10^{-4} \, \mbox{GeV}^{-1} \; ,
\label{W_bound_mu} \\
|\rho_{(\tau)}^0| &\lesssim & 1.1 \times 10^{-3} \, \mbox{GeV}^{-1} \; .
\label{W_bound_tau}
\end{eqnarray}
\end{subequations}


\subsection{The $Z$ decay width} \label{sec_Z_decay}
\indent

As a second application we calculate the correction to the decay width of the $Z$ boson into a lepton/anti-lepton and a neutrino/anti-neutrino pair. The tree-level amplitude for $Z \rightarrow \overline{f} \, f$, with $f$ being either a lepton or a neutrino, is
\begin{equation}\label{amp_Z}
iM(Z \rightarrow \overline{f} \, f)=\epsilon_{\mu}(k) \, \overline{u}_{f}(q) \, V^\mu_{Z\bar{f} f}(k) \, v_{\bar{f}}(q')
\end{equation}
with the vertex factor $V^\mu_{Z\bar{f} f}(k)$ (including the SM and LSV terms)
\begin{eqnarray}
V^\mu_{Z \bar{f} f}(k) &=& -\frac{ig}{4\cos\theta_{W}} \, \gamma^{\mu} \left(g_{V}-\gamma_{5} \right) \nonumber \\
& + & \delta_{\ell f} \left(c_3^{[\mu} \gamma^{\nu]} + c_4^{[\mu} \gamma^{\nu]}\gamma_5 \right) k_{\nu}  \nonumber \\
& + &  \frac{1}{4} \left(1 - \delta_{\ell f}\right)  v_4^{[\nu} \gamma^{\mu]} \left(1-\gamma_5\right) k_{\nu} \; ,\label{vertex_Zff}
\end{eqnarray}
where we have briefly introduced the Kroenecker delta $\delta_{\ell f}$, which is one if $f$ is a lepton ($f = \ell$) and zero otherwise ($f = \nu_\ell$). As in the SM, $g_{V}=1 - 4\sin^2\theta_{W}$ for $f=\ell$ and $g_{V}=1$ for $f=\nu_{\ell}$.


\subsubsection{$Z$ decay into charged leptons} \label{sec_Z_decay_ll}
\indent

Let us start with the $Z$ decaying into a lepton/anti-lepton pair. The calculation is very similar to the one leading to eq.~\eqref{amp_W_2} and the unpolarized tree-level amplitude for the process is
%
\begin{equation}\label{amp_Z_ll}
\langle |M|^2\rangle \!=\! \frac{g^2 m_Z^2 \left(1\!+\!g_{V}^{2}\right) }{12\cos^2 \theta_W} \! \left[ 1 \! + \! \frac{16\cos^2 \theta_W m_Z^2 }{g^2 \left(1\!+\!g_{V}^{2}\right)} \left( c_{30}^2 \!+\! c_{40}^2 \!  \right) \right]
\end{equation}
%
and we notice that, again, no SM-LSV interference term is left, so that the first non-zero correction is of second order in the LSV parameters.

Plugging this fully isotropic amplitude into eq.~\eqref{Gamma} ({\it mutatis mutandis}), dividing by the full $Z$ width $\Gamma_Z$ and using $m_W = \cos\theta_W m_Z$ gives us the branching ratio
\begin{equation}\label{gamma_Z_ll}
{\rm BR}(Z \rightarrow \overline{\ell} \ell) \!=\! \frac{G_{F} m_{Z}^3 \left(1 \!+\! g_{V}^{2}\right) }{24\sqrt{2}\pi \Gamma_Z} \! \left[ 1 \! +\!  \frac{ 2 \sqrt{2} }{1 \!+\! g_V^2} \! \left( \frac{ c_{30}^2\!  \! + c_{40}^2}{ G_{F} } \right) \! \right].
\end{equation}

The decay rates of the $Z$ boson into lepton/anti-lepton pairs have been experimentally determined and read~\cite{PDG}
\begin{subequations}
\begin{eqnarray}
{\rm BR}(Z \rightarrow \overline{e} \, e)_{\rm exp} &=& \left( 3.3632  \pm 0.0042 \right) \, \% \; ,
\label{exp_lim_Z1_ee} \\
{\rm BR}(Z \rightarrow \overline{\mu} \, \mu)_{\rm exp} &=& \left( 3.3662 \pm 0.0066 \right) \, \% \; ,
\label{exp_lim_Z1_mumu} \\
{\rm BR}(Z \rightarrow \overline{\tau} \, \tau)_{\rm exp} &=& \left( 3.3696 \pm 0.0083 \right) \, \% \; ,
\label{exp_lim_Z1_tautau}
\end{eqnarray}
\end{subequations}
so that, assuming once again that the LSV effects are buried under the respective experimental errors, we find the following upper bounds (at $1 \, \sigma$)  
\begin{subequations}
\begin{eqnarray}
\sqrt{c_{(e)30}^2+ c_{(e)40}^2} &\lesssim&  \, 7.4  \times 10^{-5} \, \mbox{GeV}^{-1} \; ,
\label{Z_bound_1_e} \\
\sqrt{c_{(\mu)30}^2+ c_{(\mu)40}^2} &\lesssim&  \, 9.3  \times 10^{-5} \, \mbox{GeV}^{-1} \; ,
\label{Z_bound_1_mu} \\
\sqrt{c_{(\tau)30}^2+ c_{(\tau)40}^2} &\lesssim&  \, 1.1  \times 10^{-4} \, \mbox{GeV}^{-1} \; .
\label{Z_bound_1_tau}
\end{eqnarray}
\end{subequations}

\subsubsection{$Z$ decay into neutrinos} \label{sec_Z_decay_nunu}
\indent

Next we consider the $Z$ boson decaying into a neutrino/anti-neutrino pair. For this process we find that the unpolarized amplitude is (using $g_V = 1$)
\begin{equation}\label{amp_Z_nunu}
\langle |M|^2  \rangle = \frac{g^2 m_Z^2 }{6\cos^2 \theta_W} \left( 1 +  \frac{\cos^2 \theta_W m_Z^2}{g^2} v_{40}^2  \right) \; ,
\end{equation}
so that the corresponding branching ratio is
\begin{equation}\label{gamma_Z_nunu}
{\rm BR}(Z \rightarrow \overline{\nu}_\ell \, \nu_\ell) = \frac{G_{F} \, m_{Z}^3  }{12\sqrt{2}\pi \; \Gamma_Z} \left( 1 +  \frac{v_{40}^2}{4 \sqrt{2} \, G_F} \right) \; .
\end{equation}

In collision experiments neutrinos are not directly detected due to their feeble interactions with matter. Detectors in high-energy experiments usually measure the tracks of electrons, muons and photons, which are typical final products of the heavier particles emerging in energetic collisions. From these tracks it is possible to reconstruct the energy and momentum of the original products of the collision and, given that energy and momentum are conserved, it is then possible to infer how much energy and momentum are missing and these are attributed to so-called {\it invisible} products. In the SM, the three neutrino families are able to successfully account for the partial width into invisible final states~\cite{Mandl}.

The inferred branching ratio of the $Z$ boson into invisible products is~\cite{PDG, LEP}
\begin{equation} \label{exp_lim_Z2}
{\rm BR}(Z \rightarrow {\rm invisible})_{\rm exp} = \left( 20.000 \pm 0.055 \right) \, \% \; ,
\end{equation}
which is well fitted by the SM assuming lepton universality, i.e., essentially using the first term in eq.~\eqref{gamma_Z_nunu} multiplied by three to account for the neutrino families.

Contrary to the two other processes already discussed, here we are unable to explicitly differentiate between the lepton families in the final states, as invisible final states cannot be further discriminated. However, from the bounds above, we see that the magnitudes of the LSV coefficients are not considerably different, varying up to a factor of two. We shall therefore assume that the LSV coefficients for the three families have comparable sizes and treat them as identical, i.e., $v_{(e)40} = v_{(\mu)40} = v_{(\tau)40} $.

Under these circumstances, from eq.~\eqref{exp_lim_Z2} we see that the relative uncertainty is~$\sim 3 \times 10^{-3}$, so we obtain the following $1 \, \sigma$ upper bound
\begin{equation}\label{Z_bound_2}
|v_{(\ell)40}| \lesssim 4 \times 10^{-4} \, \mbox{GeV}^{-1} \; ,
\end{equation}
\newline

\subsection{Muon decay} \label{sec_muon_decay}
\indent

Now we  analyse the process $\mu^- \rightarrow \nu_\mu \, e^- \, \bar{\nu}_e$, which accounts to practically 100\% of the branching ratio for muon decay; other channels are responsible for  $< 1\%$ of the total decay rate and will be ignored~\cite{PDG}. This is also a purely leptonic process and we analyse it at tree level, where the amplitude is
%
\begin{eqnarray}\label{amp_mu}
iM(\mu^{-} \rightarrow \nu_\mu e^{-} \bar{\nu}_e) & = & \frac{\eta_{\alpha \beta}}{m_W^2} \, \bar{u}_\mu (p_1) V^\alpha_{W\ell \nu_\ell}(p_1-p_3) u_{\nu_\mu}(p_3) \nonumber \\
& \times & \bar{u}_e(p_4) V^\beta_{W\ell \bar{\nu}_\ell}(p_2 + p_4) v_{\nu_e}(p_2) \; ,
\end{eqnarray}
%
with the interaction vertex given by eq.~\eqref{vertex_W}.

It is important to observe that the vertex is defined with the momentum transfer $k$ flowing into the vertex, so that the LSV part of $V^\mu_{W\ell \bar{\nu}_\ell}(k)$ has opposite signs in the two vertices. Another issue here is the family index implicit in each vertex above. The only LSV couplings involved are $\rho_{(e)}$ and $\rho_{(\mu)}$, which were constrained in eqs.~\eqref{W_bound_e} and~\eqref{W_bound_mu}, respectively. These bounds differ by only $2\%$, thus allowing us to simplify the expressions below by making $\rho_{(e)} = \rho_{(\mu)} \equiv \rho$.

Taking care of the sign of the momentum transfer and averaging over the initial spin, we obtain the following amplitude squared
\begin{eqnarray}
\langle |M|^2  \rangle &  = & \frac{1}{128 m_W^4} \! {\rm Tr} \! \left[ \overline{\Gamma}_{-\mu} \left(1 - \gamma_5 \right) \; \slashed{p}_2 \;  \left(1 + \gamma_5 \right) \overline{\Gamma}_{+\nu} \;  \slashed{p}_4 \right]  \nonumber \\
& \times & {\rm Tr} \! \left[ \overline{\Gamma}^\mu_{+} \left(1 - \gamma_5 \right) \left( \slashed{p}_1 + m_\mu \right) \left(1 + \gamma_5 \right) \overline{\Gamma}^\nu_{-}  \;\slashed{p}_3 \right]
\end{eqnarray}
where the matrix operators are defined in eq.~\eqref{v_operator}. As usual, we have neglected the mass of the electron relative to that of the muon. The squared amplitude is then the sum of the following partial amplitudes:
\begin{widetext}
\begin{eqnarray}
\langle |M|^2  \rangle_{\rm SM} & = & \frac{g^4  m_\mu^2}{m_W^4} \; E_2 (m_\mu - 2 E_2) \; ,  \label{amp_mu_SM}  \\
\langle |M|^2  \rangle_{\rm LSV}^{(1)} & = & \frac{ g^3 m_\mu^2  }{m_W^4} \; p_2^\mu p_3^\nu p_4^\alpha  \rho^\beta \epsilon_{\mu \nu \alpha \beta} \; ,  \; \label{amp_mu_LSV_1}  \\
\langle |M|^2  \rangle_{\rm LSV}^{(2)} & = & \frac{g^2 m_\mu^2}{4 m_W^4} \; \bigg\{  m_\mu^2 \rho^2 E_3 (2 E_3 - m_\mu) + 4E_2 (m_\mu -2 E_2) \left[ (p_2\cdot\rho)^2 + (p_4\cdot\rho)^2 \right] \nonumber \\
& + &   2 (m_\mu \rho_0 - p_3\cdot\rho) \left[ 2 E_2 m_\mu \rho_0 (m_\mu -2 E_2)  +  p_3\cdot\rho (4 E_2^2 - 2 (E_2 + E_3) m_\mu + m_\mu^2) \right]  \nonumber \\
& + &  2 p_2\cdot\rho \left[ m_\mu(2 E_2 - m_\mu)(m_\mu \rho_0 - p_3\cdot\rho) - 2 p_4\cdot\rho (4 E_2^2 - 2 E_2 m_\mu - E_3 m_\mu)  \right]
\nonumber \\
& - &  2 m_\mu p_4\cdot\rho (m_\mu -2 E_4) (m_\mu \rho_0 - p_3\cdot\rho)  \bigg\} \; .   \label{amp_mu_LSV_2}
\end{eqnarray}
\end{widetext}

The decay rate of the muon is the generalization of eq.~\eqref{Gamma} to the case of a three-body decay, that is
\begin{eqnarray}\label{gamma_mu_1}
\Gamma(\mu^{-} \rightarrow \nu_\mu \, e^{-} \, \bar{\nu}_e)= \frac{1}{16 (2\pi)^5 m_\mu} \int \frac{d^3{\bf p}_2 \; d^3{\bf p}_3 \; d^3{\bf p}_4}{E_2 E_3 E_4} \nonumber \\
\times \, \langle|M|^{2}\rangle \; \delta^{(4)}\left(p_1 - p_2 - p_3 - p_4 \right) ,
\hspace{0.3cm}
\end{eqnarray}
which can be simplified using $\delta^4( p_1 - p_2 - p_3 - p_4) = \delta(m_\mu - E_2 - E_3 - E_4) \, \delta^3(  {\bf p}_2 + {\bf p}_3 + {\bf p}_4 )$, thus making ${\bf p}_3 = -  \left( {\bf p}_2 + {\bf p}_4 \right)$ and $E_3 = |{\bf p}_2 + {\bf p}_4|$ upon integration over ${\bf p}_3$. This leaves us with
\begin{eqnarray}\label{gamma_mu_2}
\Gamma(\mu^{-} \rightarrow \nu_\mu \, e^{-} \, \bar{\nu}_e) = \frac{1}{16 (2\pi)^5 m_\mu} \int \frac{d^3{\bf p}_2 \; d^3{\bf p}_4}{E_2 E_3 E_4} \nonumber \\
 \times \; \langle|M|^{2}\rangle \, \delta(m_\mu - E_2 - E_3 - E_4) \; .
\end{eqnarray}

We now integrate over ${\bf p}_2$, the 3-momentum of the electron neutrino. Here we may set the $z$ axis along ${\bf p}_4$, which is constant at this point, so that $d^3{\bf p}_2 = E_2^2 dE_2 \sin\theta_2 d\theta_2 d\phi_2$. The $\theta_2$ integral may be approached using $E_3 = \sqrt{E_2^2 + E_4^2 + 2E_2 E_4 \cos\theta_2}$. With this substitution we integrate over $\theta_2$ and eq.~\eqref{gamma_mu_2} reduces to
\begin{eqnarray}\label{gamma_mu_3}
\Gamma(\mu^{-} \rightarrow \nu_\mu \, e^{-} \, \bar{\nu}_e) = \frac{1}{16 (2\pi)^5 m_\mu} \int \frac{d^3{\bf p}_4}{E_4^2} \nonumber \\
 \times \; \int d\phi_2 \, dE_2 \, \langle|M|^{2}\rangle \; ,
\end{eqnarray}
where the $\phi_2$ and $E_2$ integrals must be evaluated under the following conditions: ${\bf p}_4 = E_4 \hat{{\bf z}}$, $E_3 = m_\mu - E_2 - E_4$ and $\cos\theta_2 = \left( E_3^2 - E_2^2 - E_4^2  \right)/2 E_2 E_4$.

Unlike the SM case, where at this point only ${\bf p}_2$ and ${\bf p}_4$ must be considered, we have also the background $\rho^\mu$ contracted to the outgoing momenta. It is then convenient to write ${\bm \rho} =|{\bm \rho}| (\sin \theta_\rho \cos \phi_\rho, \; \sin \theta_\rho \sin\phi_\rho, \; \cos \theta_\rho)$ and ${\bf p}_2 = E_2(\sin \theta_2 \cos \phi_2, \; \sin \theta_2 \sin \phi_2, \;\cos \theta_2)$, so that the integral over $\phi_2$ may be performed. We will not quote this intermediate result explicitly, but we remark that, due to the totally anti-symmetric contractions, the first-order amplitude (cf.~eq.~\eqref{amp_mu_LSV_1}) vanishes identically.

\begin{figure}[t!]
\begin{minipage}[c]{1.09\linewidth}
\includegraphics[width=\linewidth]{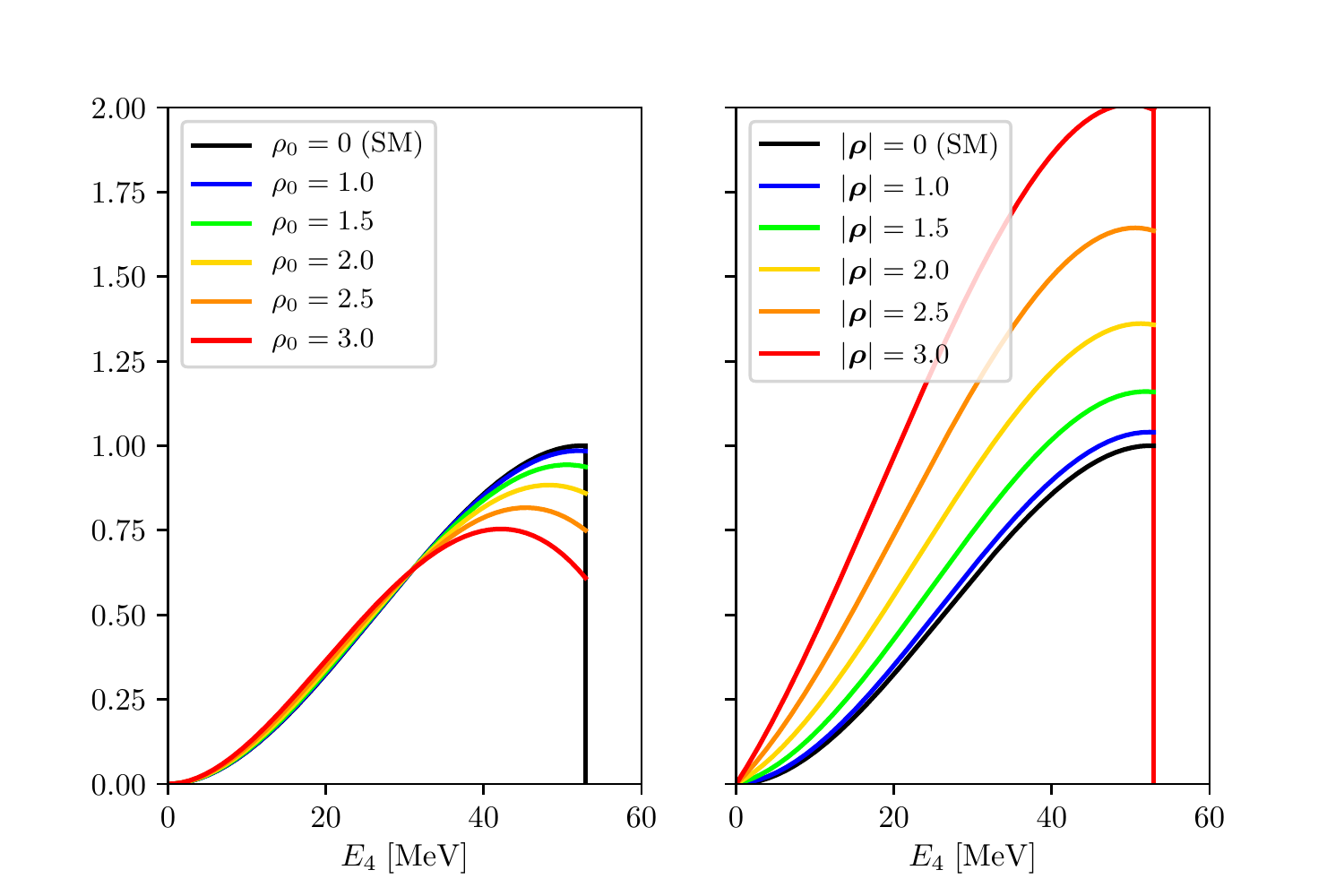}
\end{minipage} \hfill
\caption{Normalized energy spectrum of the emitted electrons (cf.~eq.~\eqref{spectrum}) for different values of the LSV parameter (in units of ${\rm GeV}^{-1}$). Left panel: $\rho_0 \neq 0$ and ${\bm \rho} = 0$; right panel: $\rho_0 = 0$ and ${\bm \rho} \neq 0$. Here we made $\rho_{(e)} = \rho_{(\mu)} \equiv \rho$. }
\label{fig_spectrum}
\end{figure}

The limit of the $E_2$ integral is determined by the kinematics to be $E_2 = \left[ m_\mu/2, m_\mu/2 - E_4\right]$. After performing this integral, the only dynamic variable is $p_4$, the 4-momentum of the electron, which appears in combination with the LSV background. The same trick as above may be employed here, i.e., we let ${\bm \rho} = |{\bm \rho}|\hat{{\bf z}}$ and write ${\bf p}_4 = E_4(\sin \theta_4 \cos \phi_4, \; \sin \theta_4 \sin \phi_4, \;\cos \theta_4)$, so that $d^3{\bf p}_4 = E_4^2 dE_4 d\Omega_4$. After integrating over $\Omega_4$ we obtain the energy spectrum of the emitted electrons with the LSV correction (making $t \equiv E_4/m_\mu$ and denoting the decay rate by $\Gamma_\mu$ for short)
\begin{eqnarray} \label{spectrum}
\frac{d\Gamma_\mu}{dE_4} & \!=\! & \frac{g^4 m_\mu^4 t^2 (3 - 4t)}{384\pi^3 m_W^4} \! \bigg\{ 1\! - \! \frac{ \left(8t^3 - 50t^2 + 65t - 15 \right) \! m_\mu^2 \rho_0^2 }{10 g^2 (3 - 4t)}  \nonumber \\
& \!-\! & \frac{ \left(8t^4 - 8t^3 + 39t^2 + 10t - 30 \right) \! m_\mu^2 |{\bm \rho}|^2 }{20 t g^2 (3 - 4t)}  \bigg\} \; ,
\end{eqnarray}
which is shown in fig.~\eqref{fig_spectrum} for different -- unrealistically large -- values of the LSV parameters.

Equation~\eqref{spectrum} displays a few interesting features. For the SM (at tree level) the peak energy of the emitted electron is $E_4^{\rm max} = m_\mu/2$, which is also a kinematical threshold imposed by momentum conservation. The inclusion of LSV disturbs the general shape of the spectrum as shown in fig.~\eqref{fig_spectrum}, where we see that a purely time-like background would suppress the spectrum, whereas a purely space-like $\rho$ would enhance it. The peak energy also recedes from its LSV-free value at different rates for purely time- or space-like components. All these effects could potentially be searched for in sensitive experiments, specially if time-stamped data are taken (see discussion in section~\ref{sec_conclusion} and in appendix~\ref{ap_SCF}).

Finally, integrating eq.~\eqref{spectrum} in the range $E_4 = \left[ 0, m_\mu/2 \right]$ we obtain the decay rate of the muon
\begin{eqnarray}\label{gamma_mu}
\Gamma_\mu = \frac{ G_F^2 \, m_\mu^5}{192 \pi^3} \left[  1 + \frac{113 m_\mu^2}{15360 m_W^2 G_F} \left(  \rho_0^2 + \frac{55 |{\bm \rho}|^2}{226} \right) \right]
\end{eqnarray}
whose first term is the tree-level result from the SM. The experimentally measured lifetime of the muon is~\cite{PDG}
\begin{equation}\label{exp_lim_mu}
\tau_\mu = \Gamma_\mu^{-1} = (2.1969811 \pm 0.0000022)\times 10^{-6} \, {\rm s} \; ,
\end{equation}
and, by demanding that the second term in eq.~\eqref{gamma_mu} be smaller than the relative uncertainty from eq.~\eqref{exp_lim_mu} ($\sim~10^{-6}$), we find the following bound at the $1 \, \sigma$ level
\begin{equation} \label{muon_bound}
\sqrt{\rho_0^2 + \frac{55 |{\bm \rho}|^2}{226}}  \lesssim 3 \times 10^{-2} \, \mbox{GeV}^{-1} \; .
\end{equation}

\section{Concluding remarks} \label{sec_conclusion}
\indent

We studied a modification to the Glashow-Salam-Weinberg electroweak model through non-minimal couplings in the non-Abelian and Abelian sectors of the lepton-boson interaction. These couplings introduce LSV via two (family-dependent) real 4-vectors that give rise to a preferred orientation in space-time. Our results show that such LSV interactions would lead to modifications in the branching ratios of the $W$ and $Z$ bosons, as well as to the lifetime of the muon.

The respective amplitudes have been evaluated at tree level and we found that, for all processes considered, the LSV parameters only contribute to second order. The SM-LSV interference terms drop out from the amplitudes for $W^{-} \rightarrow \ell \, \overline{\nu}_{\ell}$ and $Z \rightarrow \overline{\nu}_\ell \, \nu_\ell$ due to anti-symmetry (cf. eq.~\eqref{amp_W_LSV_1}) or, in the case of $Z \rightarrow \overline{\ell} \, \ell$ and $\mu^- \rightarrow \nu_\mu \, e^- \, \bar{\nu}_e$, they automatically cancel in the squared amplitude. This is in line with the results from a number of analyses of scattering and decay processes~\cite{Maluf, scat_LSV, SouzaPLB, YJ, Castro, erratumUFMA}.

Using recent experimental results we are able to constrain the magnitude of combinations of the (family-dependent) LSV parameters, cf. eqs.~\eqref{W_bound_e}-\eqref{W_bound_tau}, \eqref{Z_bound_1_e}-\eqref{Z_bound_1_tau}, \eqref{Z_bound_2} and~\eqref{muon_bound}. It is important to note that these bounds were obtained in the rest frame of the decaying particles, but he LSV parameters are not static as seen from the particle's own rest frame, not to mention from Earth's rotating reference frame. Therefore, we need to introduce a reference frame in which the LSV tensors are -- at least approximately -- static. A convenient option is the so-called Sun-centered frame (SCF), which is discussed in appendix~\ref{ap_SCF}.

The measurements determining the $W$ and $Z$ widths (and branching ratios) have a center-of-mass energy $\sim$~100~GeV, which is the same order of magnitude of their masses~\cite{OPAL_W, OPAL_Z}, so that the respective Lorentz factors $\gamma_{\rm rest}$ are very close to unity ($\beta \ll 1$). The MuLan experiment~\cite{MuLan} used muons created through pion decay with momenta $\sim 30$ MeV, which also amounts to very small Lorentz factors. Therefore, the components of a generic LSV 4-vector $V^\mu$ in the laboratory frame (LAB) are approximately equal to those in the rest frame, i.e. $V^\mu_{\rm LAB} \approx \gamma_{\rm rest} V^\mu_{\rm rest}$, where factors proportional to $\gamma_{\rm rest} \beta$ may be neglected. With $\gamma_{\rm rest} \approx 1$ and using eqs.~\eqref{V_T} and~\eqref{V_I} after averaging over $T_\oplus$, we have
\begin{eqnarray}
\left( V^{0}_{\rm rest} \right)^2 & \approx &  \left( V^{T}_{\rm SCF} \right)^2 \; ,  \label{V2_T} \\
|{\bf V}_{\rm rest}|^2 & \approx &  \left( \frac{1}{2} + s_{\chi}^2  \right) \left(V^{X}_{\rm SCF}\right)^2 + \left(  \frac{1}{2} + c_{\chi}^2  \right)\left(V^{Y}_{\rm SCF}\right)^2
\nonumber \\
& + & \left(V^{Z}_{\rm SCF}\right)^2 - 2c_{\chi}s_{\chi}\left(V^{X}_{\rm SCF}\right)\left(V^{Y}_{\rm SCF}\right) \; .  \label{V2_I}
\end{eqnarray}
so we can translate our bounds -- obtained in the rest frame of the decaying particles -- into the SCF.

As previously noted, the bounds for the different families do not differ significantly -- by at most a factor of 1.5 (see eqs.~\eqref{Z_bound_1_e} and~\eqref{Z_bound_1_tau}) -- so we shall adopt a simplifying approach and use the overall branching ratios for $W^{-} \rightarrow \ell \; \overline{\nu}_{\ell}$ and $Z \rightarrow \overline{\ell} \, \ell$ from ref.~\cite{PDG}, which read
\begin{eqnarray}
{\rm BR}(W^{-} \rightarrow \ell \, \overline{\nu}_{\ell})_{\rm exp} = \left( 10.86 \pm 0.09 \right) \, \% \; , \label{br_W_ell}   \\
{\rm BR}(Z \rightarrow \overline{\ell} \, \ell)_{\rm exp} = \left( 3.3658 \pm 0.0023 \right) \, \% \; , \label{br_Z_ell}
\end{eqnarray}
thus allowing us to obtain the following $1 \, \sigma$ upper bounds
\begin{eqnarray}
|\rho^0_{(\ell)}| \lesssim 8 \times 10^{-4} \, \mbox{GeV}^{-1} \; , \label{bound_W_generic}   \\
\sqrt{c_{(\ell)30}^2+ c_{(\ell)40}^2} \lesssim  \, 5  \times 10^{-5} \, \mbox{GeV}^{-1} \; . \label{bound_Z_generic}
\end{eqnarray}

Now we are finally able to translate our local limits into the SCF for a generic lepton family $(\ell)$. By using eqs.~\eqref{Z_bound_2}, \eqref{bound_W_generic} and~\eqref{bound_Z_generic} the bounds on combinations of the time-like components in the SCF read 
\begin{widetext}
\begin{eqnarray}
|\rho^T_{\rm (\ell) SCF}| & \lesssim & 8 \times 10^{-4} \, \mbox{GeV}^{-1} \; , \label{W_bound_SCF} \\
\sqrt{\left( \rho^T_{\rm (\ell) SCF} \right)^2 + 0.6 \left( \xi^T_{\rm (\ell) SCF} \right)^2 +  \left( \rho^T_{\rm (\ell) SCF} \right) \left( \xi^T_{\rm (\ell) SCF} \right)} & \lesssim & 2 \times 10^{-4} \, \mbox{GeV}^{-1} \; ,  \label{Z_1_bound_SCF} \\
|\rho^T_{\rm (\ell) SCF} - 0.6\xi^T_{\rm (\ell) SCF}| & \lesssim & 5 \times 10^{-4} \, \mbox{GeV}^{-1} \; ,  \label{Z_2_bound_SCF}
\end{eqnarray}
\end{widetext}
whereas from eqs.~\eqref{muon_bound} with eqs.~\eqref{V2_T} and~\eqref{V2_I}, we find
\begin{widetext}
\begin{equation}
\sqrt{\left( \rho^T_{\rm (\ell) SCF} \right)^2 + 0.23 \left(\rho^{X}_{\rm (\ell) SCF}\right)^2 + 0.25 \left(\rho^{Y}_{\rm (\ell) SCF}\right)^2 +0.24  \left(\rho^{Z}_{\rm (\ell) SCF}\right)^2 - 0.24 \left(\rho^{X}_{\rm (\ell) SCF}\right)\left(\rho^{Y}_{\rm (\ell) SCF}\right)} \lesssim 3 \times 10^{-2} \, \mbox{GeV}^{-1} \; ,  \label{mu_bound_SCF}
\end{equation}
\end{widetext}
where we used $\sin^2\theta_W = 0.23$ and $\chi \approx 43^\circ$ for the co-latitude of the MuLan experiment in Villigen, Switzerland~\cite{MuLan}. We note in passing that the limits above do not constrain the spatial components of the 4-vector $\xi$.

\begin{figure}[t!]
\begin{minipage}[b]{1.09\linewidth}
\includegraphics[width=\linewidth]{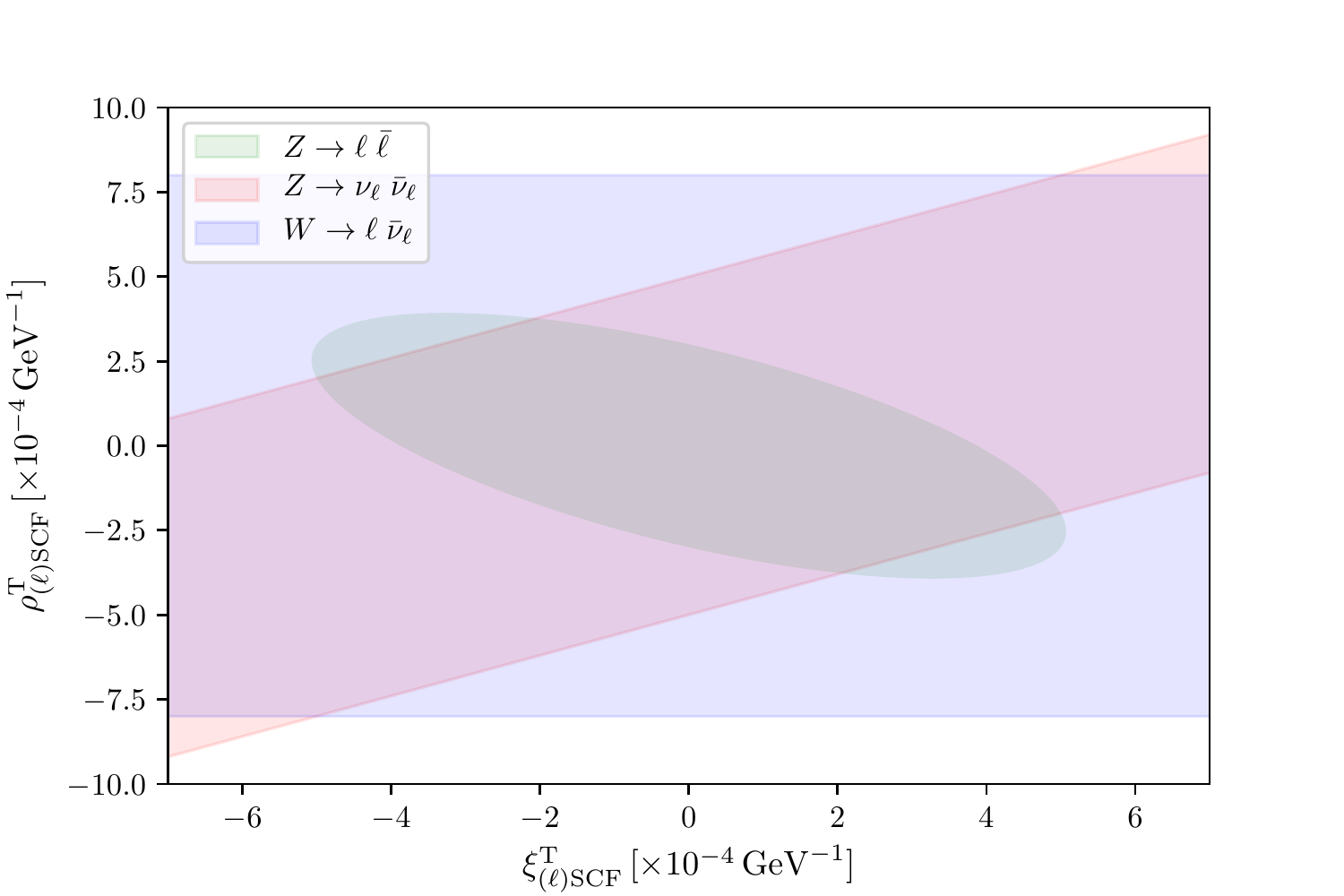}
\end{minipage} \hfill
\caption{Allowed regions for time components of the LSV parameters in the SCF from eqs.~\eqref{W_bound_SCF}, \eqref{Z_1_bound_SCF} and \eqref{Z_2_bound_SCF}. The curves represent 1~$\sigma$ limits and do not differentiate lepton families, as the family-specific bounds do not differ significantly.}
\label{fig_bounds}
\end{figure}

In fig.~\ref{fig_bounds} we show the allowed regions for the time components of the LSV backgrounds from eqs.~\eqref{W_bound_SCF}, \eqref{Z_1_bound_SCF} and~\eqref{Z_2_bound_SCF} without discriminating among lepton families, as the family-specific bounds are similar. Incidentally, the coupling studied in ref.~\cite{scat_LSV} in the context of QED is analogous to $c_1$ in eq.~\eqref{eqfinal}, but our bounds are about one order of magnitude stronger.

In table~\ref{table:list_vert_sm} we list LSV corrections to existing SM vertices, whereas in table~\ref{table:list_vert_lsv} we present terms originally not possible at tree level in the SM. Of particular interest is the $\gamma \; \nu \; \bar{\nu}$ term, which endows the neutrino with a tree-level electromagnetic interaction that could be detected as a magnetic or electric dipole moment -- both only possible in the SM at loop level and including non-zero neutrino masses~\cite{Schrock, Agostini}. Also the quartic couplings in table~\ref{table:list_vert_lsv} would provide distinctive signatures of LSV in collider experiments, especially $W^{+} \, W^{-} \, \ell \, \bar{\ell}$ and $W^{+} \, W^{-} \, \nu_{\ell} \, \bar{\nu}_{\ell}$. These represent LSV-induced vector-boson fusion as contact interactions, thus strongly contrasting with the loop-mediated SM processes~\cite{ATLAS}.


As a closing remark we note that similar non-minimal couplings were proposed in ref.~\cite{UFMA}. The main difference is that we introduce LSV in the $SU_{\rm L}(2)$ and $U(1)_{\rm Y}$ sectors simultaneously. This is important after spontaneous symmetry breaking and becomes particularly evident in $Z$ decays, where both $\xi$ and $\rho$ contribute to the amplitudes, cf. eqs.~\eqref{amp_Z_ll} and \eqref{amp_Z_nunu}. Furthermore, in ref.~\cite{UFMA} the authors report a first-order LSV correction to the amplitudes. This is not the case, as shown here and in other works on LSV non-minimal couplings in scattering processes~\cite{Maluf, scat_LSV, SouzaPLB, YJ, Castro}. We have pointed this out and the authors issued an erratum to their original paper~\cite{erratumUFMA}. Their limits are nonetheless compatible with ours modulo small multiplicative factors of order one.

\begin{acknowledgments}
The authors are grateful to J. A. Hela\"yel-Neto for interesting discussions. We also thank the anonymous referee for relevant criticism. This work was funded by the Brazilian National Council for Scientific and Technological Development (CNPq). P.C.M. expresses his gratitude to COSMO - CBPF for the hospitality. The work of M.J.N. was supported by CNPq under grant 313467/2018-8 (GM) and he also thanks the University of Alabama for the hospitality.
\end{acknowledgments}

\appendix

\section{Sun-centered frame for LSV} \label{ap_SCF}
\indent

In LSV models Lorentz symmetry is broken through tensors that transform differently under observer- and particle-Lorentz transformations and that are fixed in space-time, i.e., they are static backgrounds. This means that there is a reference frame where the LSV 4-vectors are fixed, but the physical observables that we have discussed are measured in Earth-bound reference frames and as such cannot be taken as static. For this reason we must look for a convenient reference frame where the aforementioned coefficients are fixed.

It is clear that a frame fixed to Earth's surface will not suffice, as it is a non-inertial reference frame, so we cannot expect an external background to be fixed from our point of view -- in fact we would see it rotating. The next -- and perhaps most convenient -- possibility is to use a reference frame fixed relative to the Sun. This is a good choice for a few reasons: it is approximately inertial over the time scale of most experiments (its motion around the galaxy has a period of $\sim 200$ million years), it is experimentally accessible, and may have its axes conveniently oriented relative to the Earth.

We will adopt the Sun-centered frame (SCF) as a standard reference frame where the LSV coefficients are time independent~\cite{tables}. Therefore, relative to an observer fixed on Earth, the background will seem to rotate, so experimental signals affected by LSV should generally present time modulations, specially with sidereal frequencies. In fact, even isotropic backgrounds in the SCF will appear to be anisotropic in our frame because of both rotational and translational motions of the Earth, which produce boosts. In this sense, rotation-invariance violations are a key signal for Lorentz violations in Earth-bound experiments (also in space-based tests~\cite{sat}).

According to refs.~\cite{tables, sat}, the axes in the SCF are defined such that the $Z$ axis is directed parallel to Earth's rotational axis, $X$ points from the Sun to the vernal equinox, while $Y$ completes a right-handed system; the origin of time $T$ is at the 2000 vernal equinox. Regarding the standard Earth-bound frame for a point in the northern hemisphere, the $z$ axis is vertical from the surface (points to the local zenith), $x$ points south and $y$ points east. The local time $T_{\oplus}$ is defined to be the time measured in the SCF from one of the moments when $y$ lies along $Y$.


To see how we can make the passage from the LSV coefficients in the laboratory frame (LAB), where they are in general time dependent, to the SCF, where they are fixed, we use a generic background $V^{\mu}$. The components of this vector in the two frames are connected via
\begin{equation}\label{SCF_1}
V^{\mu}_{\rm LAB} = \Lambda^{\mu}_{\,\,\,\ \nu} \, V^{\nu}_{\rm SCF} \; ,
\end{equation}
with $\Lambda^{\mu}_{\,\,\,\ \nu}$ representing an observer Lorentz transformation between Earth and the SCF. From now on, we represent the components of $V^{\mu}$ in the LAB frame by $V^{\rm 0,x,y,z}_{\rm LAB}$ and those in the SCF by $V^{\rm T,X,Y,Z}_{\rm SCF}$.

The explicit form of the (time-dependent) Lorentz transformation $\Lambda^{\mu}_{\,\,\,\ \nu}$ is
%
\begin{equation}\label{SCF_transf}
\Lambda^{0}_{\,\,T} = 1, \quad \Lambda^{0}_{\,\,I} = -{\bm \beta}^{I}, \quad \Lambda^{i}_{\,\,T} = - (R\cdot{\bm \beta})^{i}, \quad \Lambda^{i}_{\,\,I} = R^{iI} \; ,
\end{equation}
where ${\bm \beta}$ is the velocity (${\bf v}/c$ in natural units) of the LAB relative to the SCF and $R^{iJ}$ is a spatial rotation. Notice that the Lorentz factor $\gamma = 1/ \sqrt{1 - \beta^2}$ is essentially unity due to the smallness of the relative speed of Earth relative to the Sun.

The boost components are given by ($\eta \approx  23.4^{\circ}$ is the inclination of Earth's axis relative to the orbital plane)
\begin{eqnarray}
{\bm \beta}^X & = &  \beta_{\oplus} \, \sin(\Omega_{\oplus}T) - \beta_{L} \, \sin(\omega_{\oplus}T_{\oplus}) \; , \\
{\bm \beta}^Y & = &   -\beta_{\oplus} \, \cos\eta \, \cos(\Omega_{\oplus}T) + \beta_{L} \, \cos(\omega_{\oplus}T_{\oplus}) \; ,  \\
{\bm \beta}^Z & = &   -\beta_{\oplus} \, \sin\eta \, \cos(\Omega_{\oplus}T)
\end{eqnarray}
and, defining $\sin\chi \equiv s_{\chi}$, $\cos\chi \equiv c_{\chi}$; $\sin(\omega_{\oplus}T_{\oplus}) \equiv s_{\oplus}$, $\cos(\omega_{\oplus}T_{\oplus}) \equiv c_{\oplus}$, the matrix $R^{iJ}$ is given by
\begin{equation}\label{rot_matrix}
R^{iJ} = \left(
\begin{array}{ccc}
c_{\chi} \, c_{\oplus} & c_{\chi} \, s_{\oplus} & -s_{\chi} \\
-s_{\chi} & c_{\chi} & 0 \\
s_{\chi} \, c_{\oplus} & s_{\chi} \, s_{\oplus} & c_{\chi} \\
\end{array}
\right) \; .
\end{equation}

The $\Lambda^{i}_{\,\,T} = - (R\cdot{\bm \beta})^{i}$ read
\begin{eqnarray}
\Lambda^{x}_{\,\,T} & = & -c_{\chi} \, c_{\oplus} \, \beta^X -c_{\chi} \, s_{\oplus} \, \beta^Y + s_{\chi} \, \beta^Z \label{L_xT} \; ,   \\
\Lambda^{y}_{\,\,T} & = & s_{\chi} \, \beta^X - c_{\chi} \, \beta^Y \label{L_yT} \; ,  \\
\Lambda^{z}_{\,\,T} & = & - s_{\chi} \, c_{\oplus} \, \beta^X - s_{\chi} \, s_{\oplus} \, \beta^Y - c_{\chi} \, \beta^Z \; , \label{L_zT}
\end{eqnarray}
where the numerical values of the parameters appearing above are
\begin{eqnarray*}
\beta_{\oplus} & \approx & 10^{-4} , \; \textmd{Earth's orb. vel.} \\
\beta_{L} & = & r_{\oplus} \, \omega_{\oplus} \, \sin\chi <  10^{-6} , \; \textmd{Earth's rot. vel.} \\
\omega_{\oplus} & = & 2\pi/\textmd{day} \approx 7 \times 10^{-5} \,  {\rm s}^{-1} , \; \textmd{Earth's rot. angular vel.}  \\
\Omega_{\oplus} & = & 2\pi/\textmd{year} \approx 2 \times 10^{-7} \, {\rm s}^{-1} , \; \textmd{Earth's orb. angular vel.} \\
\chi & = & \textrm{experiment's co-latitude} \; .
\end{eqnarray*}

From the values above we see that $\Lambda^{0}_{\,\,I} = -{\bm \beta}^{I}$ and $\Lambda^{i}_{\,\,T} = - (R\cdot{\bm \beta})^{i}$ are suppressed due to the smallness of the boost factors and may be safely ignored. Applying this to our generic vector we find that its components are translated from the LAB frame to the SCF as
\begin{eqnarray}
V^{0}_{\rm LAB} & = &  V^{\rm T}_{\rm SCF} + \mathcal{O}(\beta) \; , \label{V_T} \\
V^{i}_{\rm LAB} & = &  R^{iI}V^{\rm I}_{\rm SCF} + \mathcal{O}(\beta) \; , \label{V_I}
\end{eqnarray}
which means that, up to very small contributions proportional to boost factors, time and space components of $V_{\rm LAB}$ and $V_{\rm SCF}$ do not mix. We are therefore able to separately analyse LSV background 4-vectors that have either purely time or spatial components in the SCF.

\section{Connection with $a_F^{(5) \mu \alpha \beta}$ and $b_F^{(5) \mu \alpha \beta}$} \label{ap_a_F}
\indent

At the end of sec.~\ref{sec_model_lg} we mentioned that the first terms in eq.~\eqref{eqfinal} are analogous to $a_{\rm F}^{(5) \mu\alpha\beta}$ and $b_{\rm F}^{(5) \mu\alpha\beta}$, which couple the vector and pseudo-vector currents, respectively, to the photon field-strength tensor~\cite{tables}. Indeed, following Table P58 we have
\begin{eqnarray}
a_F^{(5) \mu\alpha\beta} & \rightarrow & -\frac{1}{2} F_{\alpha\beta} \bar{\psi} \gamma_\mu \psi   \; , \\
b_F^{(5) \mu\alpha\beta} & \rightarrow & -\frac{1}{2} F_{\alpha\beta} \bar{\psi} \gamma_5 \gamma_\mu \psi \; .
\end{eqnarray}

The couplings $a_F^{(5) \mu\alpha\beta}$ may be treated as rank-3 tensors. These are anti-symmetric in $(\alpha, \beta)$ because of the field-strength tensor, meaning that there are in total $6 \times 4 = 24$ components. As such they may be decomposed into
\begin{equation}
a_F^{(5) \mu \alpha \beta} = \frac{1}{3} (\eta^{\mu \alpha} t^\beta - \eta^{\mu \beta} t^\alpha) + \frac{1}{6}\varepsilon^{\mu \alpha \beta \kappa} (t')_\kappa + \bar{a}^{\mu \alpha \beta} \, ,
\end{equation}
with the different terms satisfying
\begin{eqnarray}
\eta_{\alpha\gamma}\bar{a}^{\alpha \beta \gamma} & = & \varepsilon_{\mu \alpha \beta \kappa} \bar{a}^{\alpha \beta \kappa} = 0 \; , \\
\eta_{\alpha\gamma}a_F^{(5)\alpha \beta \gamma} & = & t^\beta   \; , \\
\varepsilon_{\mu \alpha \beta \kappa} a_F^{(5)\alpha \beta \kappa} & = & t'_\mu   \; .
\end{eqnarray}

The LSV Lagrangian with the $a_F^{(5)}$ term (and similarly for $b_F^{(5)}$) may be written using the irreducible components as
\begin{eqnarray}
\mathcal{L}_{\rm LSV}^{d=5} & \supset & -\frac{1}{2} a_F^{(5) \mu \alpha \beta} \bar{\psi} \gamma_\mu \psi F_{\alpha\beta}  \\
& &\hspace{-1.2cm} =\frac{1}{3}  t^\alpha  \bar{\psi} \gamma^\beta \psi F_{\alpha \beta} + \frac{1}{6} (t')^\alpha \bar{\psi} \gamma^\beta \psi \tilde{F}_{\alpha \beta}  - \frac{1}{2}\bar{a}^{\mu \alpha \beta}  \bar{\psi} \gamma_\mu \psi F_{\alpha \beta} \, , \nonumber
\end{eqnarray}
where $\tilde{F}_{\alpha \beta} = \frac{1}{2}\varepsilon_{\alpha \beta \mu \kappa}F^{\mu \kappa}$. The first term above is compatible with the first two terms in eq.~\eqref{eqfinal} and we see that (in the charged lepton sector) the vector and pseudo-vector coefficients may be expressed as
\begin{eqnarray}
c_1^\beta & = & \frac{1}{3}\eta_{\alpha\gamma}a_{F,\ell}^{(5)\alpha \beta \gamma}  \; , \\
c_2^\beta & = & \frac{1}{3}\eta_{\alpha\gamma}b_{F,\ell}^{(5)\alpha \beta \gamma} \; ,
\end{eqnarray}
where the sub index $\ell$ refers to the lepton sector.

Similarly, looking at the neutrino sector, an analogous correspondence can be found, namely
\begin{equation}
\frac{1}{4}v_2^\beta = \frac{1}{3}\eta_{\alpha\gamma}a_{F, \nu}^{(5)\alpha \beta \gamma} = -\frac{1}{3}\eta_{\alpha\gamma}b_{F,\nu}^{(5)\alpha \beta \gamma}  \; ,
\end{equation}
where the sub index $\nu$ refers to the neutrino sector. This indicates that $a_F^{(5)}$ and $b_F^{(5)}$, despite being essentially combinations of $\xi_{(\ell)}$ and $\rho_{(\ell)}$, will also be different for charged and neutral leptons in each family.


\end{document}